\documentclass[a4paper]{PoS}

\usepackage{amsmath}
\usepackage{cite}
\usepackage{subcaption}

\newcommand{\be}{\begin{equation}}
\newcommand{\ee}{\end{equation}}
\newcommand{\bea}{\begin{eqnarray}}
\newcommand{\eea}{\end{eqnarray}}

\title{Isospin-breaking corrections to the muon magnetic anomaly in Lattice QCD}

\ShortTitle{Isospin-breaking corrections to $a_\mu^{HVP}$}

\author{\speaker{D.~Giusti}$^{(a,b)}$, V.~Lubicz$^{(a,b)}$, G.~Martinelli$^{(c)}$, F.~Sanfilippo$^{(b)}$, and S.~Simula$^{(b)}$

\\

\it $^{(a)}$ Dipartimento di Matematica e Fisica, Universit\`a  degli Studi Roma Tre, Rome, Italy.\\ Email: \email{davide.giusti@uniroma3.it}, \email{vittorio.lubicz@uniroma3.it}

\it $^{(b)}$ Istituto Nazionale di Fisica Nucleare, Sezione di Roma Tre, Rome, Italy.\\ Email: \email{francesco.sanfilippo@infn.it}, \email{simula@roma3.infn.it}

\it $^{(c)}$ Dipartimento di Fisica, Universit\`a  degli Studi di Roma "La Sapienza" and INFN, Sezione di Roma, Rome, Italy.\\ Email: \email{martinelli@roma1.infn.it}

}

\abstract{In this contribution we present a lattice calculation of the leading-order electromagnetic and strong isospin-breaking (IB) corrections to the quark-connected hadronic-vacuum-polarization (HVP) contribution to the anomalous magnetic moment of the muon.
The results are obtained adopting the RM123 approach in the quenched-QED approximation and using the QCD gauge configurations generated by the ETM Collaboration with $N_f = 2+1+1$ dynamical quarks, at three values of the lattice spacing ($a \simeq 0.062, 0.082, 0.089$ fm), at several lattice volumes and with pion masses between $\simeq 210$ and $\simeq 450$ MeV.
After the extrapolations to the physical pion mass and to the continuum and infinite-volume limits the contributions of the light, strange and charm quarks are respectively equal to $\delta a_\mu^{\rm HVP}(ud) = 7.1 ~ (2.5) \cdot 10^{-10}$, $\delta a_\mu^{\rm HVP}(s) = -0.0053 ~ (33) \cdot 10^{-10}$ and $\delta a_\mu^{\rm HVP}(c) = 0.0182 ~ (36) \cdot 10^{-10}$.
At leading order in $\alpha_{em}$ and $(m_d - m_u) / \Lambda_{QCD}$ we obtain $\delta a_\mu^{\rm HVP}(udsc) = 7.1 ~ (2.9) \cdot 10^{-10}$, which is currently the most accurate determination of the IB corrections to $a_\mu^{\rm HVP}$.}

\FullConference{The 9th International workshop on Chiral Dynamics\\
		17-21 September 2018\\
		Durham, NC, USA}

\begin{document}

\section{Introduction}
\label{sec:intro}

The anomalous magnetic dipole moments of charged leptons $a_\ell$ are defined as the deviations of the spin gyromagnetic ratios $g_\ell$ from the result predicted by the Dirac equation, $a_\ell = (g_\ell - 2) / 2$.
Leptonic magnetic anomalies arise in quantum field theories as a result of virtual loop fluctuations.
In this respect, they can be viewed as windows to quantum loops including effects due to new degrees of freedom beyond the Standard Model (SM) of Particle Physics.

In the case of the muon, $a_\mu$ is one of the most accurately determined dimensionless physical quantity in Nature: it is currently known both experimentally~\cite{Bennett:2006fi} and from a SM theoretical calculation~\cite{PDG} to approximately $0.5$ ppm.
Intriguingly, there is a long-standing discrepancy between the BNL E821 experimental value and the SM prediction at the $3 \sigma \div 4 \sigma$ level.
Since this tension might be an exciting indication of New Physics beyond the SM, an intense research program is currently underway in order to achieve a significant reduction of the experimental and theoretical uncertainties.
New $(g - 2)$ experiments at Fermilab (E989)~\cite{Logashenko:2015xab} and J-PARC (E34)~\cite{Otani:2015lra} aim at a fourfold reduction of the experimental uncertainty such that a similar reduction in the theoretical uncertainty is of timely interest.
Hadronic loop contributions due to the HVP and hadronic light-by-light terms~\cite{Jegerlehner:2017gek} give rise to the main theoretical uncertainty and, with a view to the planned experimental accuracy, they will soon become a major limitation of this SM test.

Nowadays the theoretical predictions for the hadronic contribution $a_\mu^{\rm HVP}$ are most accurately determined using dispersion relations for relating the HVP function to the experimental cross section data for $e^+ e^-$ annihilation into hadrons~\cite{Davier:2010nc,Hagiwara:2011af}.
However, since the pioneering works of Refs.~\cite{Lautrup:1971jf,deRafael:1993za,Blum:2002ii}, lattice QCD calculations of $a_\mu^{\rm HVP}$ (see Ref.~\cite{Miura:2019xtd} for a recent review) have been made an impressive progress providing a completely independent cross-check from first principles.

With the increasing accuracy of lattice calculations, it becomes mandatory to include electromagnetic (em) and strong IB corrections, which contribute to the HVP to ${\cal O} (\alpha^3_{em})$ and ${\cal O} (\alpha^2_{em} (m_d - m_u)/\Lambda_{QCD})$, respectively.
Here we present the results of a lattice calculation of the IB corrections to the HVP contribution due to light-, strange- and charm-quark (connected) intermediate states, obtained in Ref.~\cite{Giusti:2019xct} using the RM123 approach~\cite{deDivitiis:2011eh,deDivitiis:2013xla}, which consists in the expansion of the path integral in powers of the mass difference ($m_d - m_u$) and of the em coupling $\alpha_{em}$.
The quenched-QED (qQED) approximation, which treats the dynamical quarks as electrically neutral particles, has been adopted and quark-disconnected contractions have not been included yet because of the large statistical fluctuations of the corresponding signals.

\section{Isospin-breaking corrections in the RM123 approach}
\label{sec:IB}

We have evaluated the HVP contribution $a^{\rm HVP}_\mu$ to the muon $(g - 2)$ by adopting the time-momentum representation ~\cite{Bernecker:2011gh}, namely
\be
    a^{\rm HVP}_\mu = 4 \alpha_{em}^2 \int_0^\infty dt ~ K_\mu(t) V(t) ~ ,
    \label{eq:amu_t}
\ee
where the kernel function $K_\mu(t)$ is given by
\be
    K_\mu(t) = \frac{4}{m_\mu^2} \int_0^\infty d\omega ~ \frac{1}{\sqrt{4 + \omega^2}} ~ 
                      \left( \frac{\sqrt{4 + \omega^2} - \omega}{\sqrt{4 + \omega^2} + \omega} \right)^2 
                      \left[ \frac{\mbox{cos}(\omega m_\mu t) - 1} {\omega^2} + \frac{1}{2} m_\mu^2 t^2 \right] 
    \label{eq:kernel}
\ee
with $m_\mu$ being the muon mass.
In Eq.~(\ref{eq:amu_t}) the quantity $V(t)$ is the vector current-current Euclidean correlator defined as
\be
    V(t) \equiv -\frac{1}{3} \sum_{i=1,2,3} \int d\vec{x} ~ \langle J_i(\vec{x}, t) J_i(0) \rangle ~ ,
    \label{eq:VV}
\ee
where
 \be
     J_\mu(x) \equiv \sum_{f = u, d, s, c, ...} J_\mu^f(x) =  \sum_{f = u, d, s, c, ...} q_f ~ \overline{\psi}_f(x) \gamma_\mu \psi_f(x)
     \label{eq:Jem}
 \ee
is the em current with $q_f$ being the electric charge of the quark with flavor $f$ in units of the electron charge $e$, while $\langle ... \rangle$ means the average of the $T$-product over gluon and quark fields.

We consider only the quark-connected HVP contributions, thus neglecting off-diagonal flavor terms.
In this case each quark flavor $f$ contributes separately
\be
    a^{\rm HVP}_\mu = \sum_{f = u, d, s, c, ...} [a^{\rm HVP}_\mu(f)]_{(conn)} ~ .
    \label{eq:amuf}
\ee
For sake of simplicity we drop the suffix $(conn)$, but it is understood that in the following we refer always to quark-connected contractions only.

In the RM123 method of Refs.~\cite{deDivitiis:2011eh,deDivitiis:2013xla} the vector correlator for the quark flavor $f$, $V_f(t)$, is expanded into a lowest-order contribution $V^{(0)}_f(t)$, evaluated in isospin-symmetric QCD ({\it i.e.}~$m_u = m_d$ and $\alpha_{em}=0$), and a correction $\delta V_f(t)$ computed to leading order in the small parameters $(m_d - m_u) / \Lambda_{QCD}$ and $\alpha_{em}$:
\be
     V_f(t) = V^{(0)}_f(t) + \delta V_f(t) + \dots ~ ,
\ee
where the ellipses stand for higher order terms in $(m_d - m_u) / \Lambda_{QCD}$ and $\alpha_{em}$. 

The separation between the isosymmetric QCD and the IB contributions, $V^{(0)}_f(t)$ and $\delta V_f(t)$, is prescription dependent. 
As in Ref.~\cite{Giusti:2019xct}, here we impose the matching condition in which the renormalized coupling and quark masses in the full theory, $\alpha_s$ and $m_f$, and in isosymmetric QCD, $\alpha_s^{(0)}$ and $m_f^{(0)}$, coincide in the $\overline{\rm MS}$ scheme at a scale of $2~\mbox{GeV}$.
Such a prescription is known as the Gasser-Rusetsky-Scimemi (GRS) one~\cite{Gasser:2003hk}.

\begin{figure}[htb!]
\begin{center}
\includegraphics[scale=1.5]{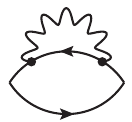} ~ 
\includegraphics[scale=1.5]{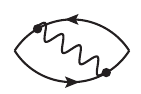} ~ 
\includegraphics[scale=1.5]{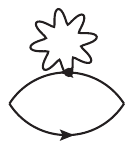} ~ 
\includegraphics[scale=1.5]{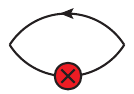} ~ 
\includegraphics[scale=1.5]{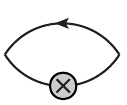}\\
~ (a) \qquad \qquad ~~ (b) \qquad \qquad ~~ (c) \qquad \qquad ~~ (d) \qquad \qquad ~~ (e) ~
\end{center}
\vspace{-0.5cm}
\caption{\it \footnotesize Fermionic connected diagrams contributing to the IB corrections $\delta a^{\rm HVP}_\mu(f)$: self-energy (a), exchange (b), tadpole (c), pseudoscalar (d) and scalar (e) insertions. Solid lines represent the propagators of the quark with flavor $f$ in isosymmetric QCD.}
\label{fig:diagrams}
\end{figure}

The calculation of the IB correlator $\delta V_f(t)$ requires the evaluation of the self-energy, exchange, tadpole, pseudoscalar and scalar insertion diagrams depicted in Fig.~\ref{fig:diagrams}.
More specifically, the IB corrections $\delta V_f(t)$ consists of two (prescription-dependent) contributions: the em, $\delta V_f^{QED}(t)$, and the strong IB (SIB), $\delta V_f^{SIB}(t)$, one.
Diagrams (\ref{fig:diagrams}a)-(\ref{fig:diagrams}d) contribute to the em corrections only, while the diagram (\ref{fig:diagrams}e) to both $\delta V_f^{QED}(t)$ and $\delta V_f^{SIB}(t)$.
Tadpole insertions (\ref{fig:diagrams}c) are a feature of lattice discretization and play a crucial role in order to preserve gauge invariance to ${\cal O} (\alpha_{em})$ in the expansion of the quark action~\cite{deDivitiis:2013xla}.
Since the lattice fermionic action used in this contribution includes a Wilson term, the insertions of pseudoscalar densities (\ref{fig:diagrams}d) account for regularization-specific IB effects associated with the tuning of the quark critical masses in the presence of em interactions~\cite{deDivitiis:2013xla,Giusti:2017dmp}.
In the numerical evaluation of the photon propagator the zero-mode has been removed according to the QED$_L$ prescription~\cite{Hayakawa:2008an}, {\it i.e.}~the photon field $A_\mu$ satisfies $A_\mu(k_0, \vec{k} = \vec{0}) \equiv 0$ for all $k_0$.

Within the qQED approximation and neglecting quark-disconnected diagrams, the QED correlator $\delta V_f^{QED}(t)$ is proportional to $\alpha_{em}~q_f^4$.
Instead, the SIB one $\delta V_f^{SIB}(t)$ is proportional to $q_f^2 ~ (m_f^{(0)} - m_f)$.
Since in the GRS prescription we require $m_f^{(0)} (\overline{\rm MS}, 2~\mbox{GeV}) = m_f (\overline{\rm MS}, 2~\mbox{GeV})$ for $f = (ud), s, c$, the SIB correlator at the renormalization scale $\mu = 2~\mbox{GeV}$ receives non-vanishing leading-order contributions only in the light quark sector (since $m_d = m_u = m_{ud}$).
In that case the correction $[ \delta V_{ud}^{SIB}(t) ] (\overline{\rm MS}, 2~\mbox{GeV})$ is proportional to the light-quark mass difference, whose value, $m_d - m_u = 2.38 \, (18)~\mbox{MeV}$ has been determined in Ref.~\cite{Giusti:2017dmp} at the physical pion mass in the $\overline{\rm MS}(2 ~ \rm GeV)$ scheme by using as inputs the experimental charged- and neutral-kaon masses.

The isosymmetric QCD gauge ensembles used in this contribution are the same adopted in Ref.~\cite{Giusti:2019xct}, {\it i.e.}~those generated by the European (now Extended) Twisted Mass Collaboration (ETMC) with $N_f = 2+1+1$ dynamical quarks, which include in the sea, besides two light mass-degenerate quarks, also the strange and the charm quarks with masses close to their physical values~\cite{Baron:2010bv}.
The gauge fields are simulated using the Iwasaki gluon action~\cite{Iwasaki:1985we}, while for sea quarks the Wilson Twisted Mass action~\cite{Frezzotti:2000nk} is employed.
Working at maximal twist our setup guarantees an automatic ${\cal O} (a)$-improvement~\cite{Frezzotti:2003ni}.
We consider three values of the inverse bare lattice coupling $\beta$, corresponding to lattice spacings varying from $0.089$ to $0.062 ~ \mbox{fm}$, pion masses in the range $M_\pi \simeq 220 \div 490 ~ \mbox{MeV}$ and different lattice volumes.
For earlier investigations of finite volume effects (FVEs) the ETM Collaboration had produced three dedicated ensembles, A40.20, A40.24 and A40.32, which share the same quark mass (corresponding to $M_\pi \simeq 320 ~ \mbox{MeV}$) and lattice spacing $(a \simeq 0.09 ~ \mbox{fm})$ and differ only in the lattice size $L$ $(L/a = 20 \div 32)$.
To improve such an investigation a further gauge ensemble, A40.40, has been generated at a larger value of the lattice size, $L/a = 40$.
For further details of the lattice simulations we refer the reader to the Appendix of Ref.~\cite{Giusti:2019xct}.

In our numerical simulations we have adopted the following local version of the vector current (see Eq.~(\ref{eq:Jem})):
\be
    J_\mu (x) = Z_A ~ q_f ~ \overline{\psi}_{f^\prime} (x) \gamma_\mu \psi_f (x) ~ ,
    \label{eq:localV}
\ee
where $\overline{\psi}_{f^\prime}$ and $\psi_f$ represent two quarks with the same mass, charge and flavor, but regularized with opposite values of the Wilson $r$-parameter ({\it i.e.}~$r_{f^\prime} = - r_f$).
Being at maximal twist the current~(\ref{eq:localV}) renormalizes multiplicatively with the renormalization constant (RC) $Z_A$ of the axial current.
By construction the local current~(\ref{eq:localV}) does not generate quark-disconnected diagrams.
As discussed in Ref.~\cite{Giusti:2017jof}, the properties of the kernel function $K_\mu(t)$ in Eq.~(\ref{eq:amu_t}), guarantee that the contact terms, generated in the HVP tensor by a local vector current, do not contribute to $a_\mu^{\rm HVP}$.

Since we have adopted the renormalized vector current (\ref{eq:localV}), the em correlator $\delta V_f^{QED}(t)$ receives a contribution from the em corrections to the RC of the vector current of Eq.~(\ref{eq:localV}) as well, namely
\be
    Z_A = Z_A^{(0)} \, \left( 1 + \frac{\alpha_{em}}{4 \pi} ~ {\cal Z}_A \right) + {\cal{O}}(\alpha_{em}^m \alpha_s^n) ~ , \qquad (m > 1, ~ n \geq 0)
    \label{eq:ZAem}
\ee
where $Z_A^{(0)}$ is the RC of the axial current in pure QCD (determined in Ref.~\cite{Carrasco:2014cwa}), while the product $Z_A^{(0)} \, {\cal{Z}}_A$ encodes  the corrections to first order in $\alpha_{em}$. 
The quantity ${\cal{Z}}_A$ can be written as
\be
    {\cal Z}_A = {\cal Z}_A^{(1)} \cdot Z_A^{fact} ~ ,
    \label{eq:ZA}
\ee
where ${\cal Z}_A^{(1)} = - 15.7963 ~ q_f^2$ is the pure QED correction to leading order in $\alpha_{em}$ given by~\cite{Martinelli:1982mw,Aoki:1998ar} and $Z_A^{fact}$ takes into account QCD corrections of order ${\cal{O}}(\alpha_s^n)$ with $n \geq 1$ to Eq.~(\ref{eq:ZA}).
It represents the QCD correction to the ``naive factorization'' approximation ${\cal{Z}}_A = {\cal{Z}}_A^{(1)}$ ({\it i.e.}~$Z_A^{fact} = 1$) adopted in Ref.~\cite{Giusti:2017jof}.
We make use of the non-perturbative determination performed in Ref.~\cite{DiCarlo:2019thl} within the RI$^\prime$-MOM scheme.

Similarly, the em corrections to the mass RC $Z_m$ enter in $\delta V_f^{QED}(t)$.
For our maximally twisted-mass setup $1 / Z_m = Z_P$ and $Z_P^{(0)}$ is the RC of the pseudoscalar density evaluated in Ref.~\cite{Carrasco:2014cwa} in isosymmetric QCD, in the $\overline{\rm MS}(2 ~ \rm GeV)$ scheme.
The pure QED contribution ${\cal{Z}}_P^{(1)} = q_f^2 ~ \left[ 6 ~ \mbox{ln}(a \mu) \right.$ $\left.- 22.5954 \right]$ to leading order in $\alpha_{em}$ is given in the $\overline{\rm MS}$ scheme at the renormalization scale $\mu$ by~\cite{Martinelli:1982mw,Aoki:1998ar}.
The values adopted for the coefficients $Z_P^{fact}$ and $Z_A^{fact}$ are collected in Table~V of Ref.~\cite{Giusti:2019xct}.

\section{Results}
\label{sec:results}

A convenient procedure~\cite{Giusti:2017jof,Giusti:2018mdh} relies on splitting Eq.~(\ref{eq:amu_t}) into two contributions corresponding to $0 \leq t \leq T_{data}$ and $t > T_{data}$, respectively.
In the first contribution the vector correlator is numerically evaluated on the lattice, while for the second contribution an analytic representation is required.
If $T_{data}$ is large enough that the ground-state contribution is dominant for $t > T_{data}$ and smaller than $T / 2$ in order to avoid backward signals, the IB corrections $\delta a_\mu^{\rm HVP}(f)$ for the quark flavor $f$ can be written as
\be
    \delta a_\mu^{\rm HVP} (f) \equiv \delta a_\mu^{\rm HVP}(<) + \delta a_\mu^{\rm HVP}(>) 
    \label{eq:deltamu}
\ee
with
\bea 
    \label{eq:deltamu_cuts_I}
    \delta a_\mu^{\rm HVP}(<) & = & 4 \alpha_{em}^2 \sum_{t = 0}^{T_{data}} K_\mu(t) ~ \delta V_f(t) ~ , \\[2mm]
    \label{eq:deltamu_cuts_II}
    \delta a_\mu^{\rm HVP}(>) & = & 4 \alpha_{em}^2 \sum_{t = T_{data} + a}^\infty K_\mu(t) ~ 
                                                          \frac{Z^{f}_V} {2 M^{f}_V} e^{- M^{f}_V t} \left[ \frac{\delta Z^{f}_V}{Z^{f}_V}
                                                         - \frac{\delta M^{f}_V}{M^{f}_V} (1 + M^{f}_V t) \right]  ~ , ~                                                   
\eea
where $M_V^{f}$ is the ground-state mass of the lowest-order correlator $V^{(0)}_f(t)$ and $Z_V^{f}$ is the squared matrix element of the vector current between the ground-state $| V_f \rangle$ and the vacuum: $Z_V^{f} \equiv (1/3) \sum_{i=x,y,z}$ $q_f^2$ $| \langle 0 | \overline{\psi}_f(0) \gamma_i \psi_f(0) | V_f \rangle |^2$.
In Refs.~\cite{Giusti:2017jof,Giusti:2018mdh} the ground-state masses $M_V^{f}$ and the matrix elements $Z_V^{f}$ have been determined for $f = (ud), s, c$ using appropriate time intervals $t_{min} \leq t \leq t_{max}$ for each value of $\beta$ and of the lattice volume for the ETMC gauge ensembles adopted in this contribution.

\begin{figure}[htb!]
\begin{center}
\includegraphics[scale=0.34]{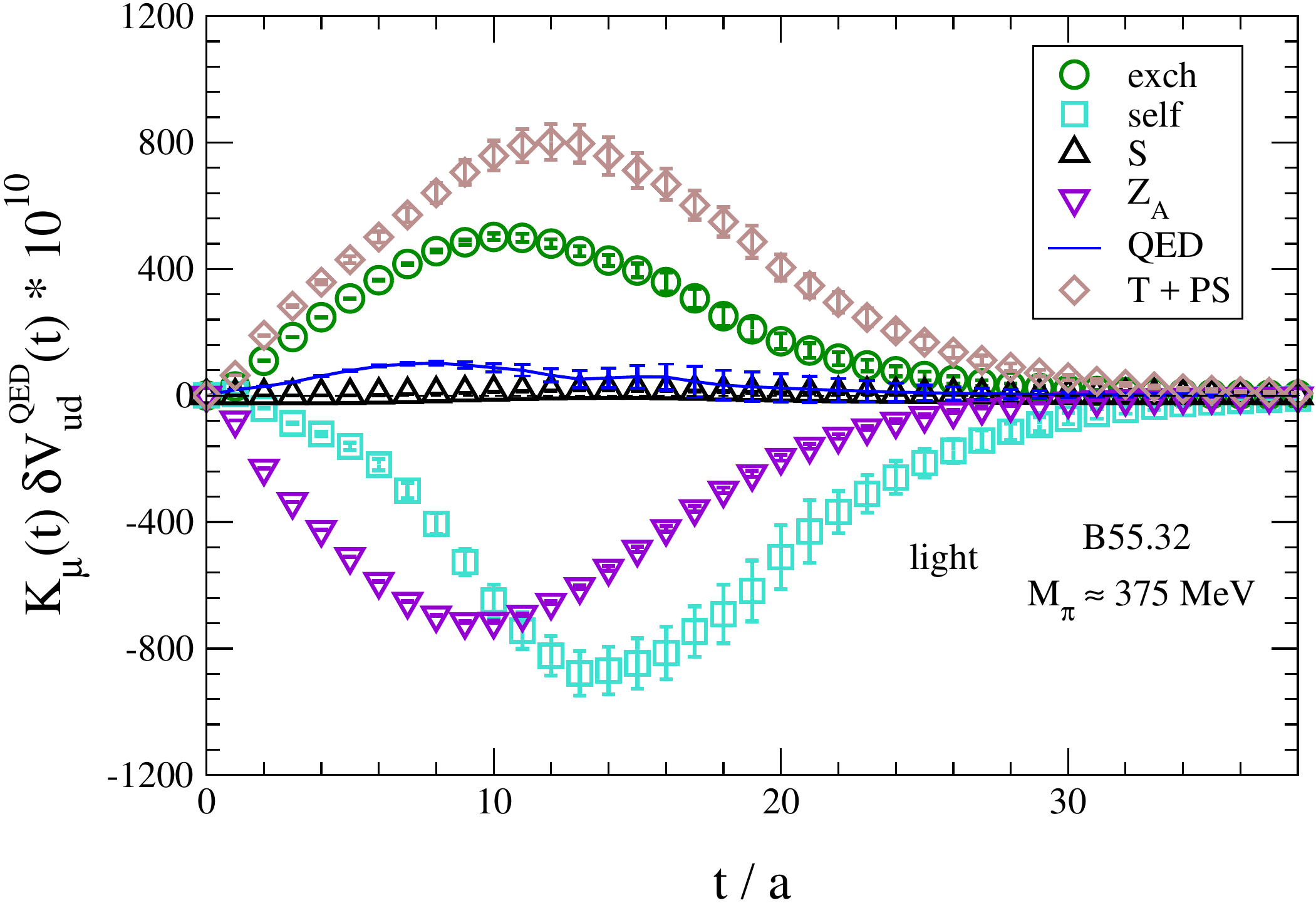} ~~ \includegraphics[scale=0.34]{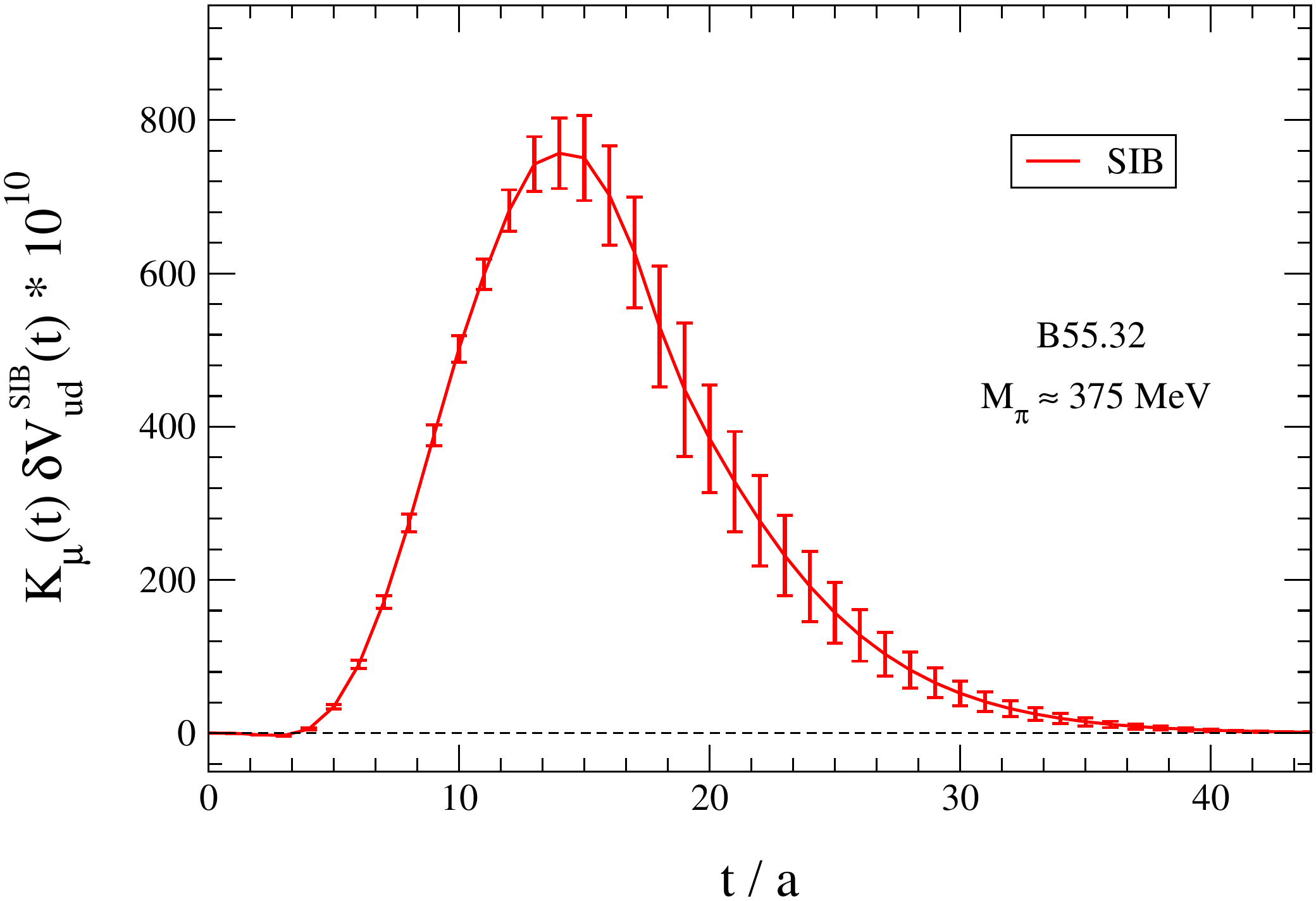}\\
\includegraphics[scale=0.34]{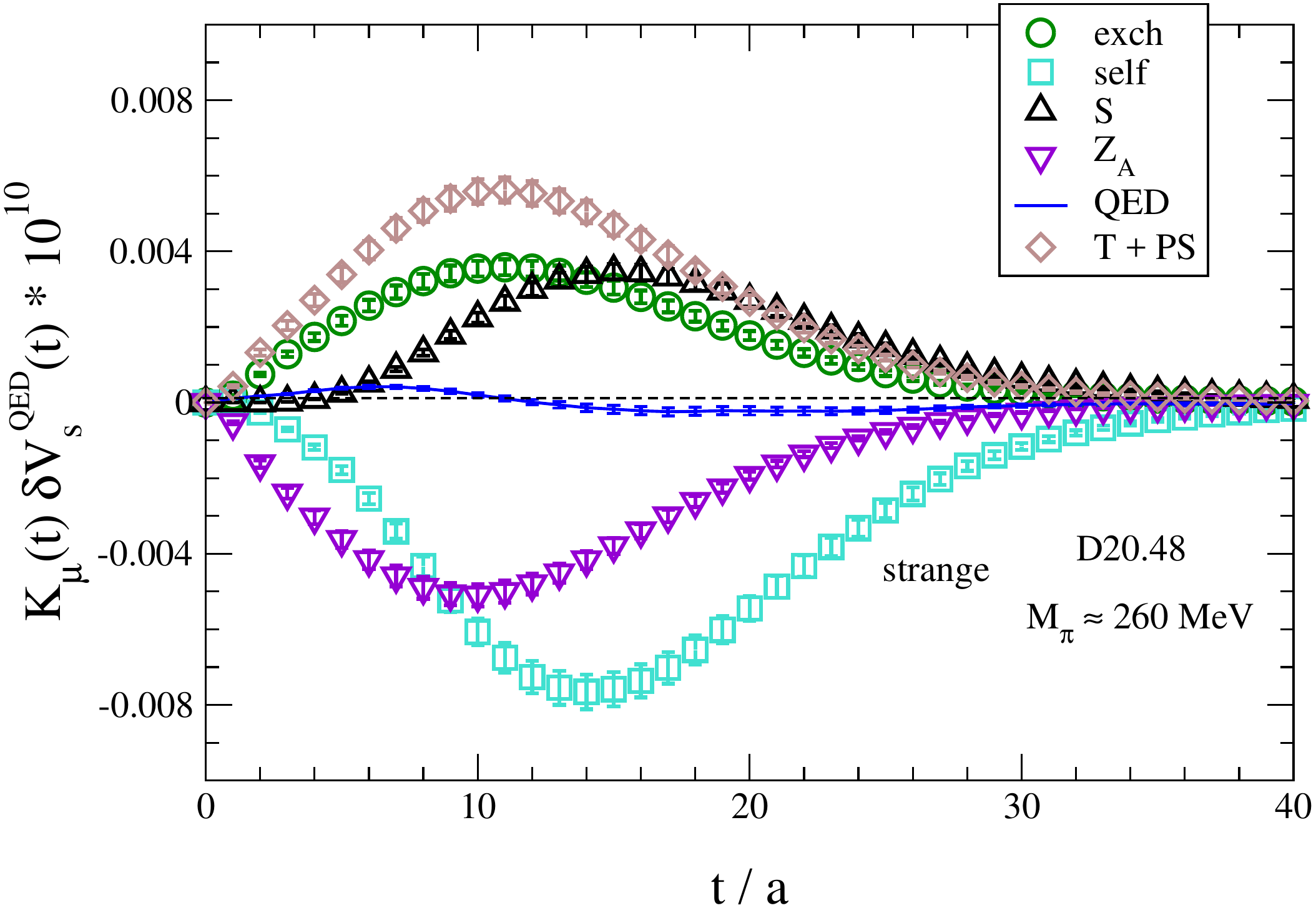} ~~ \includegraphics[scale=0.34]{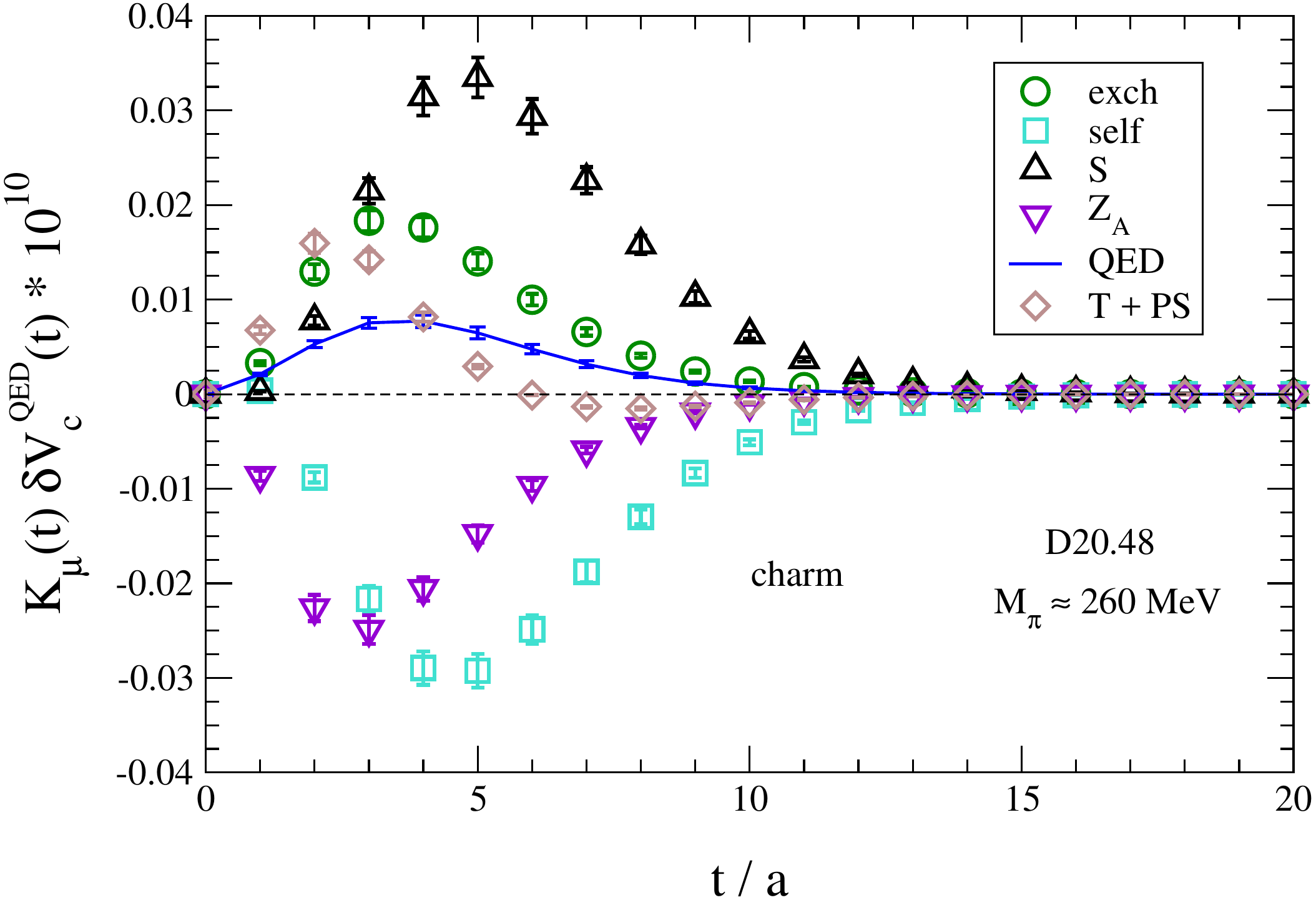}
\end{center}
\vspace{-0.5cm}
\caption{\it \footnotesize Time dependence of the integrand functions $K_\mu(t) \, \delta V^{SIB}_{ud}(t)$ (top-right panel) and $K_\mu(t) \, \delta V^{QED}_{f}(t)$ for the light- (top-left panel), strange- (bottom-left panel) and charm-quark (bottom-right panel) contributions to the IB corrections $\delta a_\mu^{\rm HVP} (f)$ [see Eq.~(\ref{eq:deltamu_cuts_I})] in the cases of  the ETMC gauge ensembles $B55.32$ ($M_\pi \simeq 375$ MeV, $a \simeq 0.082$ fm) and $D20.48$ ($M_\pi \simeq 260$ MeV, $a \simeq 0.062$ fm). In the panels the labels ``self'', ``exch'', ``T+PS'', ``S', ``${\cal Z}_A$'' indicate the QED contributions of the diagrams (\ref{fig:diagrams}a), (\ref{fig:diagrams}b), (\ref{fig:diagrams}c)+(\ref{fig:diagrams}d), (\ref{fig:diagrams}e) and the one generated by the QED corrections to the RC of the local vector current.}
\label{fig:damut}
\end{figure}

The quantities $\delta M_V^{f}$ and $\delta Z_V^{f}$ in Eq.~(\ref{eq:deltamu_cuts_II}) can be extracted respectively from the ``slope'' and the ``intercept'' of the ratio $\delta V_f(t) / V^{(0)}_f(t)$ at large time distances (see Refs.~\cite{deDivitiis:2011eh,deDivitiis:2013xla,Giusti:2017dmp,Giusti:2017jof,Giusti:2019xct}).
We have checked that the sum of the two terms in the r.h.s. of Eq.~(\ref{eq:deltamu}) is independent of the specific choice of the value of $T_{data}$ within the statistical uncertainties~\cite{Giusti:2017jof,Giusti:2019xct}.
The time dependences of the integrand functions $K_\mu(t) \, \delta V^{QED}_{f}(t)$ for $f = (ud), s, c$ and $K_\mu(t) \, \delta V^{SIB}_{ud}(t)$ are shown in Fig.~\ref{fig:damut} in the cases of the ETMC gauge ensembles $B55.32$ and $D20.48$.
After summation over the time distance $t$, the SIB contribution dominates over the QED one.

\begin{figure}[htb!]
\begin{center}
\includegraphics[scale=0.45]{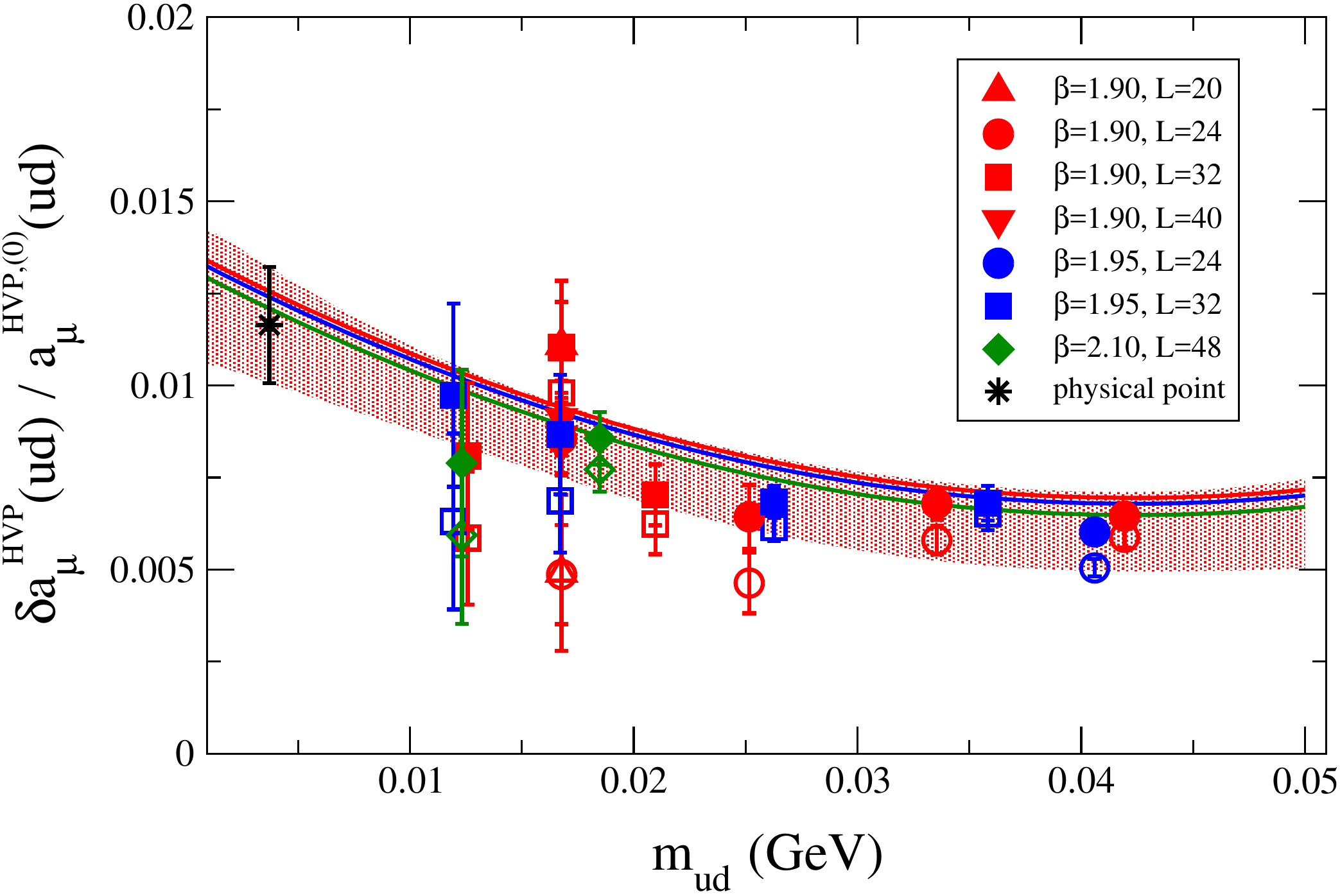}
\end{center}
\vspace{-0.5cm}
\caption{\it \footnotesize Results for the ratio $\delta a_\mu^{\rm HVP} (ud) / a_\mu^{\rm HVP,(0)} (ud)$ versus the renormalized average $u/d$ mass $m_{ud}$ in the $\overline{\rm MS}(2 ~ \rm GeV)$ scheme. The empty markers correspond to the raw data, while the full ones represent the lattice data corrected by the FVEs obtained in the fitting procedure~(\ref{eq:fit_ansatz}) with $\delta A_{1\ell}^{\ell} = 0$ and $\delta A_2^{\ell} \neq 0$. The solid lines correspond to the results of the combined fit~(\ref{eq:fit_ansatz}) obtained in the infinite-volume limit at each value of the lattice spacing. The black asterisk represents the value of the ratio $\delta a_\mu^{\rm HVP} (ud) / a_\mu^{\rm HVP,(0)} (ud)$ extrapolated to the physical pion mass, corresponding to $m^{phys}_{ud} (\overline{\rm MS}, 2 ~ \rm GeV) = 3.70 ~ (17) ~ \rm MeV$ and to the continuum and infinite-volume limits, while the red area indicates the corresponding uncertainty as a function of $m_{ud}$ at the level of one standard deviation. Errors are statistical only.}
\label{fig:ratio_ud_IB_RM123}
\end{figure}

The accuracy of the lattice data can be improved by forming the ratio of the IB corrections $\delta a_\mu^{\rm HVP}(f)$ over the leading-order terms $a_\mu^{\rm HVP,(0)} (f)$, which is shown in the case of the light-quark contribution in Fig.~\ref{fig:ratio_ud_IB_RM123}.
The attractive feature of this ratio is to be less sensitive to some of the systematics effects, in particular to the uncertainties of the scale setting.

For the combined extrapolations to the physical pion mass and to the continuum and infinite-volume limits we have adopted the following fit ansatz:
\be
    \frac{\delta a_\mu^{\rm HVP} (ud)}{a_\mu^{\rm HVP,(0)} (ud)} = \delta A_0^{\ell} \left [ 1 + \delta A_1^{\ell} \, m_{ud} + \delta A_{1\ell}^{\ell} \, m_{ud} \, 
                                                                                                         \mbox{ln}(m_{ud}) + \delta A_2^{\ell} \, m_{ud}^2 + \delta D^{\ell} \, a^2 + \delta_{FVE}^{\ell} \right] ~ ,
    \label{eq:fit_ansatz}
\ee
where the FVE term is estimated by using alternatively one of the fitting functions (see later on)
\bea
     \delta_{FVE}^{\ell} & = & \delta F^{\ell} ~ e^{- \overline{M} L} ~ \qquad \qquad \qquad \qquad \mbox{or} ~ \nonumber \\[2mm]
     \delta_{FVE}^{\ell} & = & \delta \widehat{F}^{\ell}_n ~ \frac{\overline{M}^2}{16 \pi^2 f_0^2} ~ \frac{e^{-\overline{M} L}}{(\overline{M} L)^n} ~
                                             \quad \qquad ({\rm n =} \frac{1}{2}, ~ 1, ~ \frac{3}{2}, ~ 2) ~
     \label{eq:ansatz_FVE}
\eea
with $B_0$ and $f_0$ being the leading-order low-energy constants of Chiral Perturbation Theory (ChPT) and $\overline{M}^2 \equiv 2 B_0 m_{ud}$.
For the chiral extrapolation we consider either a quadratic ($\delta A_{1\ell}^{\ell} = 0$ and $\delta A_2^{\ell} \neq 0$) or a logarithmic ($\delta A_{1\ell}^{\ell} \neq 0$ and $\delta A_2^{\ell} = 0$) dependence.
Half of the difference of the corresponding results extrapolated to the physical pion mass is used to estimate the systematic uncertainty due to the chiral extrapolation.
Discretization effects play a minor role and, for our ${\cal O} (a)$-improved simulation setup, they can be estimated by including $(\delta D^{\ell} \neq 0)$ or excluding $(\delta D^{\ell} = 0)$ the term proportional to $a^2$ in Eq.~(\ref{eq:fit_ansatz}).
The free parameters to be determined by the fitting procedure are $\delta A_0^{\ell}$, $\delta A_1^{\ell}$, $\delta A_{1\ell}^{\ell}$ (or $\delta A_2^{\ell}$), $\delta D^{\ell}$ and $\delta F^{\ell}$ (or $\delta \widehat{F}^{\ell}_n$).

Before discussing the result of the fitting procedure we focus more on the FVEs and comment on the choice of the fitting functions of Eqs.~(\ref{eq:ansatz_FVE}).
For the separate QED and SIB contributions the FVEs differ qualitatively and quantitatively, as shown in Fig.~6 of Ref.~\cite{Giusti:2019xct}.
In the case of the QED data a power-law behavior in terms of the inverse lattice size $1/ L$ is expected to start to ${\cal{O}}(1/L^3)$ because of the overall neutrality of the system~\cite{Giusti:2017jof,Giusti:2019xct,Bijnens:2019ejw}.
In the case of the SIB correlator, since a fixed value $m_d - m_u = 2.38 \, (18)$ MeV~\cite{Giusti:2017dmp} is adopted for all gauge ensembles, an exponential dependence in terms of the quantity $M_\pi L$ is expected~\cite{Lin:2001ek}.
Since the SIB contribution dominate over the QED one (see Fig.~\ref{fig:damut}), the FVEs for the ratio $\delta a_\mu^{\rm HVP} (ud) / a_\mu^{\rm HVP,(0)} (ud)$ are expected to be mainly exponentially suppressed in $M_\pi L$.~\footnote{Had we used in fitting our data~(\ref{eq:fit_ansatz}) $\delta_{FVE}^{\ell} = \delta \widetilde{F}^{\ell} / L^3$ we would have observed a change in the result~(\ref{eq:ratio_IB}) well within the statistical uncertainty.}
We remind the reader that the lowest-order term $a_\mu^{\rm HVP,(0)} (ud)$ has non-negligible FVEs, which are exponentially suppressed in terms of $M_\pi L$~\cite{Giusti:2018mdh,Lin:2001ek,Hansen:2019rbh}.
In Ref.~\cite{Giusti:2018mdh} the FVEs on $a_\mu^{\rm HVP,(0)} (ud)$ have been evaluated by using the same lattice setup adopted here and developing an analytic representation of the vector correlator based on quark-hadron duality~\cite{SVZ} at small and intermediate time distances and on the two-pion contributions in a finite box~\cite{Luscher:1990ux} at larger time distances.
After the extrapolation to the continuum limit, the lattice estimates of FVEs turn out to be much larger than the corresponding predictions of ChPT to NLO~\cite{Aubin:2015rzx}.
In Table~\ref{tab:ratio_FVEs} we collect the values of the ratio of the lattice FVEs, $\Delta_{FV}^{\rm lat} (L) \equiv a_\mu^{\rm HVP,(0)} (ud;\,L \to \infty) - a_\mu^{\rm HVP,(0)} (ud;\,L)$, computed in Ref.~\cite{Giusti:2018mdh} at the physical pion mass over the corresponding NLO ChPT predictions, $\Delta_{FV}^{\rm ChPT,NLO} (L)$.
The NNLO ChPT corrections $\Delta_{FV}^{\rm ChPT,NNLO} (L)$ have been recently computed in Ref.~\cite{Aubin:2019usy} and the ratio $\Delta_{FV}^{\rm ChPT,NNLO} (L) / \Delta_{FV}^{\rm ChPT,NLO} (L)$ for physical pion masses and lattices of size $L = 5 \div 6 ~ \mbox{fm}$ is found to be $\simeq 1.4 ~ (2)$, which points in the same direction as our lattice corrections $\Delta_{FV}^{\rm lat} (L = 5 \div 6 ~ \mbox{fm}) / \Delta_{FV}^{\rm ChPT,NLO} (L = 5 \div 6 ~ \mbox{fm}) \simeq 1.7 ~ (1)$ (see Table ~\ref{tab:ratio_FVEs}).

\begin{table}[htb!]
\begin{center}
\begin{tabular}{||c|c||c||}
\hline
$M_\pi^{phys} L$ & $L ~ (\rm fm)$ & $\Delta_{FV}^{\rm lat} (L) / \Delta_{FV}^{\rm ChPT,NLO} (L)$ \\
\hline \hline
$2.7$ & $ ~ 4.0 ~ $ & $ ~ 2.17 \, (17) ~ $ \\ 
\hline
$3.1$ & $ ~ 4.5 ~ $ & $ ~ 1.95 \, (13) ~ $ \\ 
\hline
$3.4$ & $ ~ 5.0 ~ $ & $ ~ 1.79 \, (10) ~ $ \\ 
\hline
$3.8$ & $ ~ 5.5 ~ $ & $ ~ 1.68 \, (8) ~ $ \\ 
\hline
$4.1$ & $ ~ 6.0 ~ $ & $ ~ 1.60 \, (6) ~ $ \\ 
\hline
$4.8$ & $ ~ 7.0 ~ $ & $ ~ 1.48 \, (4) ~ $ \\ 
\hline
$5.5$ & $ ~ 8.0 ~ $ & $ ~ 1.37 \, (5) ~ $ \\ 
\hline  
\end{tabular}
\end{center}
\vspace{-0.5cm}
\caption{\it \footnotesize Values of the ratio of the lattice FVEs, $\Delta_{FV}^{\rm lat} (L) \equiv a_\mu^{\rm HVP,(0)} (ud;\,L \to \infty) - a_\mu^{\rm HVP,(0)} (ud;\,L)$, computed in Ref.~\cite{Giusti:2018mdh} at the physical pion mass $M_\pi^{phys} \simeq 135 ~ {\rm MeV}$ over the NLO ChPT ones, $\Delta_{FV}^{\rm ChPT,NLO} (L)$.}
\label{tab:ratio_FVEs}
\end{table}

At the physical pion mass and in the continuum and infinite-volume limits we have obtained~\cite{Giusti:2019xct}
\be
    \frac{\delta a_\mu^{\rm HVP} (ud)}{a_\mu^{\rm HVP,(0)} (ud)} = 0.0115 ~ (18)_{stat+fit} \,  (21)_{input} \,  (20)_{chir} \, (19)_{\rm FVE}\, (9)_{a^2} \, [40] ~ ,
    \label{eq:ratio_IB}
\ee
where the errors come in the order from (statistics + fitting procedure), input parameters of the eight branches of the quark mass analysis of Ref.~\cite{Carrasco:2014cwa}, chiral extrapolation, finite-volume and discretization effects.
In Eq.~(\ref{eq:ratio_IB}) the uncertainty in the square brackets corresponds to the sum in quadrature of the statistical and systematic errors.

Using the leading-order result $a^{\rm HVP,(0)}_\mu (ud) = 619.0 ~ (17.8) \cdot 10^{-10}$ from Ref.~\cite{Giusti:2018mdh}, our determination of the leading-order IB corrections $\delta a_\mu^{\rm HVP} (ud)$ is
\be
    \delta a_\mu^{\rm HVP} (ud) =  7.1 ~ (1.1)_{stat+fit}\, (1.3)_{input}\, (1.2)_{chir}\, (1.2)_{\rm FVE}\, (0.6)_{a^2}\, [2.5] \cdot 10^{-10} ~ ,
    \label{eq:delta_ud}
\ee
which comes (within the GRS prescription) from the sum of the QED contribution
\be
    \left [ \delta a^{\rm HVP}_\mu(ud) \right ]^{(QED)} = 1.1 ~ (1.0) \cdot 10^{-10} ~ 
    \label{eq:delta_ud_QED}
\ee
and of the SIB one
\be
    \left [ \delta a^{\rm HVP}_\mu(ud) \right ]^{(SIB)} = 6.0 ~ (2.3) \cdot 10^{-10} ~ .
    \label{eq:delta_ud_SIB}
\ee
The above results show that the IB correction~(\ref{eq:delta_ud}) is dominated by the strong $SU(2)$-breaking term, which corresponds roughly to $\approx 85 \%$ of $\delta a^{\rm HVP}_\mu (ud)$.

Our determination~(\ref{eq:delta_ud}), obtained with $N_f = 2 + 1 + 1$ dynamical flavors of sea quarks, agrees within the errors with and is more precise than both the phenomenological estimate $\delta a^{\rm HVP}_\mu (ud) = 7.8 ~ (5.1) \cdot 10^{-10}$, obtained by the BMW Collaboration~\cite{Borsanyi:2017zdw} using results of the dispersive analysis of $e^+ e^-$ data~\cite{Jegerlehner:2017zsb}, and the lattice determination $\delta a^{\rm HVP}_\mu (ud) = 9.5 ~ (10.2) \cdot 10^{-10}$, obtained by the RBC/UKQCD Collaboration~\cite{Blum:2018mom} at $N_f = 2 + 1$, which includes also one disconnected QED diagram.
Recently, adopting $N_f = 1 + 1 + 1 + 1$ simulations, the FNAL/HPQCD/MILC Collaboration has found for the SIB contribution the value $\left[ \delta a^{\rm HVP}_\mu(ud) \right]^{(SIB)} = 9.0 ~ (4.5) \cdot 10^{-10}$~\cite{Chakraborty:2017tqp}.

\begin{figure}[htb!]
\begin{center}
\includegraphics[scale=0.35]{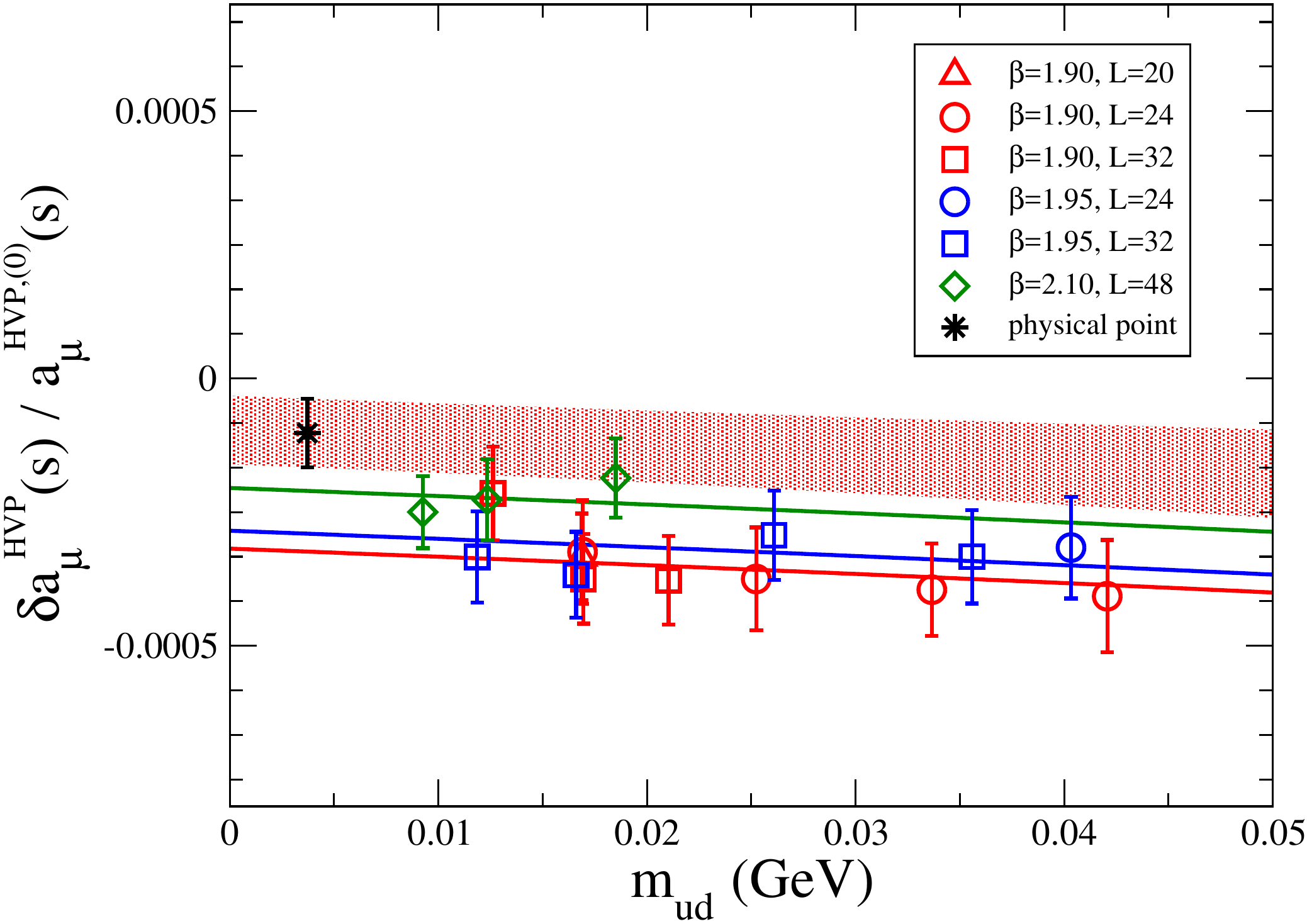} ~~ \includegraphics[scale=0.35]{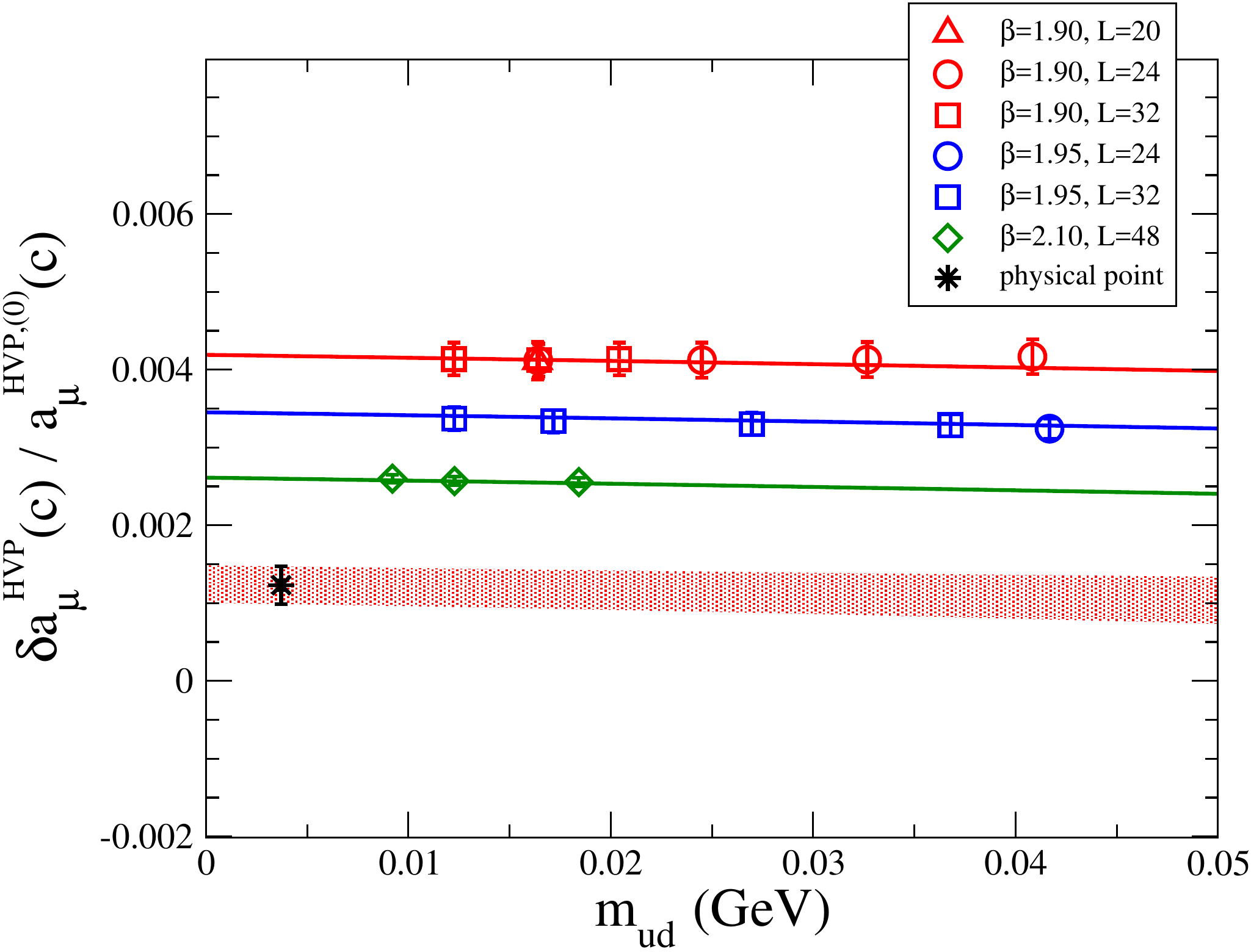}
\end{center}
\vspace{-0.5cm}
\caption{\it \footnotesize Results for the strange (left panel) and charm (right panel) contributions to $\delta a_\mu^{\rm HVP} / a_\mu^{\rm HVP,(0)} $ versus the renormalized average $u/d$ mass $m_{ud}$. The solid lines correspond to the linear fit~(\ref{eq:fit_ansatz_sc}) including the discretization term in the infinite-volume limit. The black asterisks represent the results of the extrapolation to the physical pion mass and to the continuum and infinite-volume limits while the red area indicates the corresponding uncertainty as a function of $m_{ud}$ at the level of one standard deviation. Errors are statistical only.}
\label{fig:}
\end{figure}

Thanks to the recent non-perturbative evaluation of QCD+QED effects on the RCs of bilinear operators performed in Ref.~\cite{DiCarlo:2019thl} we have updated the determinations of the strange $\delta a_\mu^{\rm HVP}(s)$ and charm $\delta a_\mu^{\rm HVP}(c)$ contributions to the IB effects made in Ref.~\cite{Giusti:2017jof}, obtaining a drastic improvement of the uncertainty by a factor of $\approx 3$ and $\approx 3.5$, respectively.
In Fig.~\ref{fig:} the updated results for the ratios $\delta a_\mu^{\rm HVP}(f) / a_\mu^{\rm HVP,(0)}(f)$ for $f = s, c$ are shown.

By adopting the same fitting function~(5.13) of Ref.~\cite{Giusti:2017jof}, namely
\be
    \frac{\delta a_\mu^{\rm HVP} (s,c)}{a_\mu^{\rm HVP,(0)} (s,c)} = \delta A_0^{s,c} + \delta A_1^{s,c} \, m_{ud} + \delta D^{s,c} \, a^2
                                                                                                          + \delta F^{s,c} \frac{1}{L^3}
    \label{eq:fit_ansatz_sc}
\ee
and after the extrapolations to the physical pion mass and to the continuum and infinite-volume limits we have found
\bea
    \label{eq:delta_s_ratio}
    \frac{\delta a_\mu^{\rm HVP}(s)}{a_\mu^{\rm HVP,(0)}(s)} & = & -0.00010 ~ (6)_{stat+fit}\, (2)_{input}\, (1)_{chir}\, (1)_{\rm FVE}\, (1)_{a^2}\, [7] \% ~ , \\[2mm]
    \label{eq:delta_c_ratio}
    \frac{\delta a_\mu^{\rm HVP}(c)}{a_\mu^{\rm HVP,(0)}(c)} & = & 0.00123 ~ (24)_{stat+fit}\, (4)_{input}\, (1)_{chir}\, (2)_{\rm FVE}\, (1)_{a^2}\, [25] \%
\eea
where the error budget has been obtained as in Ref.~\cite{Giusti:2017jof}.

Using the leading-order results $a_\mu^{\rm HVP,(0)}(s) = 53.1 ~ (2.5) \cdot 10^{-10}$ and $a_\mu^{\rm HVP,(0)}(c) = 14.75 ~ (0.56) \cdot 10^{-10}$~\cite{Giusti:2017jof}, our updated IB determinations are~\cite{Giusti:2019xct}
\bea
    \label{eq:delta_s}
    \delta a_\mu^{\rm HVP}(s) & = & -0.0053 ~ (30)_{stat+fit}\, (13)_{input}\, (2)_{chir}\, (2)_{\rm FVE}\, (1)_{a^2}\, [33] \cdot 10^{-10} ~ , \\[2mm]
    \label{eq:delta_c}
    \delta a_\mu^{\rm HVP}(c) & = & 0.0182 ~ (35)_{stat+fit}\, (5)_{input}\, (1)_{chir}\, (3)_{\rm FVE}\, (1)_{a^2}\, [36] \cdot 10^{-10}
\eea
to be compared with $\delta a_\mu^{\rm HVP}(s) = -0.018 ~ (11) \cdot 10^{-10}$ and $\delta a_\mu^{\rm HVP}(c) = -0.030 ~ (13) \cdot 10^{-10}$ given in Ref.~\cite{Giusti:2017jof}.
The updated results confirm that the em corrections $\delta a_\mu^{\rm HVP} (s)$ and $\delta a_\mu^{\rm HVP} (c)$ are negligible with respect to the current uncertainties of the corresponding lowest-order terms.
Recently~\cite{Blum:2018mom} in the case of the strange contribution the RBC/UKQCD Collaboration has found the result $\delta a_\mu^{\rm HVP} (s) = -0.0149~(32) \cdot 10^{-10}$, which deviates from our finding (\ref{eq:delta_s}) by $\approx 2$ standard deviations.

The sum of our three results (\ref{eq:delta_ud}), (\ref{eq:delta_s}) and (\ref{eq:delta_c}) yields the contribution of quark-connected diagrams to $\delta a_\mu^{\rm HVP}$ within the qQED approximation, namely $\delta a_\mu^{\rm HVP}(udsc)|_{conn} = 7.1 ~ (2.6) \cdot 10^{-10}$.
Recently, in Ref.~\cite{Blum:2018mom} one QED disconnected diagram has been calculated in the case of the $u$- and $d$-quark contribution and found to be of the same order of the corresponding QED connected term. 
Thus, we estimate that the uncertainty related to the qQED approximation and to the neglect of quark-disconnected diagrams is approximately equal to  our QED contribution (\ref{eq:delta_ud_QED}), obtaining 
\be
    \label{eq:delta_udsc}
    \delta a_\mu^{\rm HVP}(udsc) = 7.1 ~ (2.6) ~ (1.2)_{qQED+disc}\, [2.9] \cdot 10^{-10} ~ ,
\ee
which represents the most accurate determination of the IB contribution to $a_\mu^{\rm HVP}$ to date.

Using the recent ETMC determinations of the lowest-order contributions of light, strange and charm quarks, $a^{\rm HVP,(0)}_\mu (ud) = 619.0 ~ (17.8) \cdot 10^{-10}$, $a^{\rm HVP,(0)}_\mu (s) = 53.1 ~ (2.5) \cdot 10^{-10}$ and $a^{\rm HVP,(0)}_\mu (c) = 14.75 ~ (0.56) \cdot 10^{-10}$~\cite{Giusti:2017jof,Giusti:2018mdh}, and an estimate of the lowest-order quark-disconnected diagrams, $a^{\rm HVP,(0)}_\mu (disc) = -12 ~ (4) \cdot 10^{-10}$, obtained using the results of Refs.~\cite{Borsanyi:2017zdw} and~\cite{Blum:2018mom}, our finding (\ref{eq:delta_udsc}) for the IB corrections leads to an HVP contribution to the muon ($g - 2$) equal to 
\be
    a_\mu^{\rm HVP} = 682 ~ (19) \cdot 10^{-10} ~ ,
\ee
which agrees within the errors with the recent determinations based on dispersive analyses of the experimental cross section data for $e^+ e^-$ annihilation into hadrons (see, {\it e.g.}, Ref.~\cite{Davier:2019can}).

\end{document}